\documentclass[twocolumn]{jpsj3}
\def\v#1{\mib #1}
\def\dfrac#1#2{{\displaystyle\frac{#1}{#2}}}

\def\JF{J_{\rm F}}
\def\KF{K_{\rm F}}
\def\Epr{E_{\rm p}}
\def\Etw{E_{\rm tw}}
\def\Eop{E_{\rm op}}
\def\jef4{J_{\rm eff}^{(4)}}
\def\Jef2{J_{\rm eff}^{(2)}}
\def\sigtot{\v{\sigma}_{\rm tot}}
\def\tautot{\v{\tau}_{\rm tot}}
\def\simleq{\mbox{\raisebox{-1.0ex}{$\stackrel{<}{\sim}$}}}
\def\simgeq{\mbox{\raisebox{-1.0ex}{$\stackrel{>}{\sim}$}}}

\newcommand{\ket}[1]{\left\vert {#1} \right\rangle}

\def\runauthor{ Kazuo {\sc Hida}}
\title
{
Exotic Ground State Phases of $S=1/2$ Heisenberg $\Delta$-Chain with Ferromagnetic Main Chain}
\author
{
Kazuo {\sc Hida}\thanks{E-mail: hida@phy.saitama-u.ac.jp}
}

\inst
{Division of Material Science, Graduate School of Science and Engineering, Saitama University, Saitama, Saitama, 338-8570
 
}

\recdate
{
\today
}

\abst
{
The ground state phase diagram of the spin-1/2 Heisenberg frustrated $\Delta$-chain with a ferromagnetic main chain is investigated. In addition to the ferromagnetic phase, various nonmagnetic ground states are found. If the ferromagnetic coupling between apical spins and the main chain is strong, this model is approximated by a spin-1 bilinear-biquadratic chain and the spin quadrupolar phase with spin-2 gapless excitation is realized in addition to the Haldane and  ferromagnetic phases. In the regime where the coupling between the apical spins and the main chain is weak, the numerical results which suggest the possibility of a series of phase transitions among different nonmagnetic phases are obtained. Physical pictures of these phases are discussed based on the numerical results.
}

\kword
{
$\Delta$-chain, Haldane phase, spin-quadrupolar phase, Uimin-Lai-Sutherland point, ferromagnetism
}

\begin{document}
\sloppy
\maketitle
\section{Introduction} 

\begin{figure}
\centerline{\includegraphics[width=70mm]{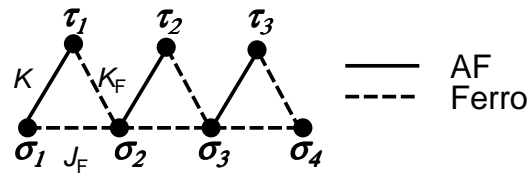}}
\caption{Structure of $\Delta$-chain.}
\label{chain}
\end{figure}

The interplay of frustration and quantum fluctuation has been extensively studied in a variety of low-dimensional quantum magnets. Even in one-dimensional cases, various  exotic quantum phenomena such as spontaneous dimerization\cite{mg}, 1/3-plateau,\cite{oku,oku2,tone,ha} and noncollinear ferrimagnetism,\cite{ym,kh} are reported. Among them, the Heisenberg $\Delta$-chain, which consists of a single main chain and apical spins interacting with the main chain as shown in Fig. \ref{chain}, has been extensively studied as one of the simplest examples of frustrated quantum spin chains.\cite{hamada,kubo,nk,sswc,tonedelta,inagaki,tend} 

In the present study, we investigate the spin 1/2 Heisenberg $\Delta$-chain with a ferromagnetic  main chain. It is assumed that one of the bonds between the apical spins and the main chain is antiferromagnetic and the other is ferromagnetic. These interactions introduce frustration. In spite of its simplicity, this model exhibits a variety of exotic quantum phases.  It should also be  noted that this model is closely related to other important theoretical models, such as a spin-1/2 ferromagnetic-antiferromagnetic alternating chain,\cite{khfa} a mixed spin chain with frustrated side chains\cite{th}, a spin-1 bilinear-biquadratic chain\cite{uimin,lai,suth,ft1,ft2,ik,lauchli}, a one-dimensional Kondo necklace\cite{doniach,sss,on,yama1,yama2,yama3} and a spin 1/2 ladder with ferromagnetic legs.\cite{km1,vjm,rm,hijii}

 In this model, the ferromagnetism of the main chain is destroyed by the frustrated coupling to the apical spins. In contrast to the mechanism of quantum destruction of antiferromagnetism, which has been extensively studied related with the high-$T_{\rm c}$ superconductivity, the mechanism of quantum destruction of {\it ferromagnetism} has been less extensively studied. Recently, however, experiments on such phenomena 
 have been reported in  one- and two-dimensional materials\cite{hase,kage} with ferromagnetic nearest neighbour and antiferromagnetic next nearest neighbour couplings. Theoretical investigation has also been carried out for corresponding models.\cite{hamada,tonehara,iq,momoi} However, we find different nonmagnetic phases  near the ferromagnetic phase in the present $\Delta$-chain.

In the next section, the model Hamiltonian is presented. Various limiting cases are discussed in \S 3. The numerical phase diagram is presented in \S 4 along with the physical description of each phase.  The last section is devoted to summary and discussion. 

\section{Hamiltonian}

We consider the spin 1/2 Heisenberg $\Delta$-chain with a ferromagnetic  main chain
 represented by a Hamiltonian

\begin{eqnarray}
{H} &=&\sum_{l=1}^{N/2} (K\v{\sigma}_{l}\v{\tau}_{l}-\JF\v{\sigma}_{l}\v{\sigma}_{l+1} - \KF\v{\sigma}_{l+1}\v{\tau}_{l}),\label{ham_delta}
\end{eqnarray}
where $K$-bonds are antiferromagnetic  and $J_{\rm F}$ and $\KF$ bonds are ferromagnetic. The operators $\v{\sigma}_l$ and $\v{\tau}_l$ are spin-$1/2$ operators. Periodic boundary condition is assumed unless specifically mentioned. In the following, we use the parametrization $k=K/\JF$ and $\alpha=\KF/K$. This model can also be regarded as a ferromagnetic-antiferromagnetic alternating chain\cite{khfa} with ferromagnetic next nearest interaction. This model is also derived as a limiting case of the mixed spin chains with frustrated side chains.\cite{th}

\section{Various Limiting Cases}
\subsection{$0 < \alpha << 1$}
For $\alpha=0$, this model can be regarded as an SU(2) symmetric version of the Kondo necklace\cite{sss,on,yama1,yama2,yama3} with a ferromagnetic main chain and antiferromagnetic Kondo interaction. Therefore, it is intuitively obvious that the singlet pairs sit on the $K$-bonds in the ground state and the ground state is gapful even for small $k$. 

 This intuition can also be  supported by the analytic results for the two-leg Heisenberg ladder with ferromagnetic legs and antiferromagnetic rungs for which the gap opens for infinitesimal rung interaction.\cite{km1,vjm,rm} In the present model with $\alpha=0$, one of the leg interactions is absent so that the gap can develop more easily with the infinitesimal $k$.
Because a finite gap cannot be destroyed by an infinitesimal perturbation, our model remains gapful for small $\alpha$.
\subsection{$0 < k << 1$}
\label{bigspin}

In this limit, the sum of the $\sigma$-spins $\v{\sigma}_{\rm tot}\equiv\sum_{l=1}^{N/2} \v{\sigma}_l$ forms a fully polarized state with $|\v{\sigma}_{\rm tot}|=N/4$. This can be regarded as a macroscopic spin with length $N/4$. Limiting the Hilbert space of $\sigma$-spins in this subspace, the Hamiltonian (\ref{ham_delta}) reduces to
\begin{eqnarray}
H_{\rm eff}&=&\frac{2}{N}\sum_{l=1}^{N/2} \left({K}\sigtot\v{\tau}_{l}- \KF\v{\sigma}_{\rm tot}\v{\tau}_{l}\right)\nonumber\\
&=&\frac{2\JF k(1-\alpha)}{N}\sigtot\tautot
\label{hameff1}
\end{eqnarray}
where $\tautot\equiv\sum_{l=1}^{N/2} \v{\tau}_l$ and constant terms are omitted. This Hamiltonian commutes with $\tautot^2$.  Therefore, the ground state has $|\sigtot+\tautot|=0$  for $0< \alpha <1$ (nonmagnetic ground state) and $|\sigtot+\tautot|=N/2$  for $\alpha >1$ (ferromagnetic ground state).

Within the effective Hamiltonian,  the lowest excitation of the effective Hamitonian (\ref{hameff1}) for $0 < \alpha <1$ is given by the state with $|\sigtot+\tautot|=1$ and $\tau_{\rm tot}=N/4$, which has the excitation energy
\begin{eqnarray}
\Delta E&=&\frac{2\JF k(1-\alpha)}{N}.
\label{upbd}
\end{eqnarray}
This vanishes in the limit $N\rightarrow\infty$. At first sight, this appears to give an upper bound for the lowest excitation energy and the present system appears to be gapless in the thermodynamic limit for sufficiently small $k$. However, this is in contradiction with the argument for small $\alpha$ in the preceding subsection.  This contradiction is resolved if the gap opens in a nonperturbative way as $\exp(-{\rm const.}/\sqrt{k})$ for small $k$ as in the case of ferromagnetic leg ladders,\cite{km1,hijii} because the above argument leading to (\ref{upbd}) is based on the perturbative argument with respect to $k$.

With the increase in $\alpha$, the ground state remains nonmagnetic until the transition to the ferromagnetic phase is reached. The behavior of the present model in this regime is, however, nontrivial. The details will be discussed using numerical methods in \S 4.

\subsection{$\alpha >> 1$} 
\label{pert} 

 In this case, the spins $\v{\sigma}_{i+1}$ and $\v{\tau}_{i}$ form a spin-1 ferromagnetic dimer.\cite{khfa} Within the lowest order approximation in $\JF$ and $K$, the effective Hamiltonian for these spin-1 operators $\hat{\v{S}}_{i}\equiv\v{\sigma}_{i+1}+\v{\tau}_{i}$ is given by
\begin{eqnarray}
H_{\rm eff}^{(1)}&=&\sum_{l=1}^{N/2} \frac{\JF}{4}\left(k-1\right)\hat{\v{S}}_{l}\hat{\v{S}}_{l+1}.
\end{eqnarray}
In the weakly frustrated regime $k >1$, the effective coupling between these ferromagnetic dimers is antiferromagnetic and the ground state is the Haldane phase in terms of $\hat{\v{S}}_{l}$'s. On the other hand, for $0 < k < 1$, the effective coupling is ferromagnetic; thus the ground state is  ferromagnetic in which all spins are aligned. Therefore, we may expect the Haldane-ferromagnetic phase transition at $k=1$. However, in the close neighbourhood of the phase boundary $k=1$, the higher order terms come into play. As a result, not only the bilinear term but also the biquadratic term appears in the effective Hamiltonian, and both become of the same order of magnitude in the vicinity of the point $k=1$ as
\begin{eqnarray}
H_{\rm eff}^{(2)}&=&\sum_{l=1}^{N/2} \left(\Jef2\hat{\v{S}}_{l}\hat{\v{S}}_{l+1}+\jef4 \left(\hat{\v{S}}_{l}\hat{\v{S}}_{l+1}\right)^2\right) \label{ham2}
\end{eqnarray}
with
\begin{eqnarray}
\Jef2 &=&\frac{\JF}{4}\left({k}-1+\frac{1}{\alpha}\right)\ , \ \ \jef4=\frac{\JF}{8\alpha}.\nonumber
\end{eqnarray}
To simplify the Hamiltonian, we set $k=1$ in the terms of $O(\alpha^{-1})$ because these terms are unimportant away from the point $k=1$. 

The exact Bethe Anszatz solution of this Hamiltonian (\ref{ham2}) is obtained for $\Jef2 =\jef4$ by Uimin\cite{uimin}, Lai,\cite{lai} and Sutherland.\cite{suth} In the following, this point is called the ULS point. At this point, the lowest  spin-2 quintuplet excitation degenerates with the lowest spin-1 triplet excitation, and both become gapless.  For $0 < \Jef2 < \jef4$, the spin-2 excitation is the lowest gapless excitation,\cite{ft1,ft2,ik} reflecting the spin-quadrupolar quasi-long range order.\cite{lauchli} This phase is called the spin-quadrupolar phase. For $ \Jef2 < 0$, the ground state becomes ferromagnetic.  Therefore, in terms of the parameters of the original Hamiltonian (\ref{ham_delta}), the ULS point is given by 
\begin{eqnarray}
k=1-\frac{1}{2\alpha}
\end{eqnarray}
and the transition to the ferromagnetic phase takes place at
\begin{eqnarray}
k=1-\frac{1}{\alpha}.
\label{percr}
\end{eqnarray}
\subsection{Stability of  Ferromagnetic Phase}

The single-magnon instability of the ferromagnetic phase  can be analytically investigated.  The Hilbert space of the state with a single magnon is spanned by the states $\ket{l, \mu}$, where $l$ is the position of the flipped spin and $\mu=\sigma$ or  $\tau$ distingushes the two types of spins. 

The Hamitonian operates on the Fourier transformed bases
\begin{eqnarray}
\ket{q,\mu}=\frac{1}{\sqrt{N/2}}\sum_l \ket{l,\mu}\exp(iql), 
\end{eqnarray}
as
\begin{eqnarray}
\Delta H\ket{q,\sigma}&=&\left(-\cos(q)+1-\frac{k(1-\alpha)}{2}\right)\ket{q,\sigma}\nonumber\\
&+&\frac{k(1-\alpha\exp(iq))}{2}\ket{q,\tau}\\
\Delta H \ket{q,\tau}&=&\frac{k(1-\alpha \exp(-iq)}{2}\ket{q,\sigma}\nonumber\\
&-&\frac{k(1-\alpha)}{2}\ket{q,\tau}
\end{eqnarray}
where $\Delta H$ is the Hamiltonian measured from the energy of the ferromagnetic state. The eigenvalue equation reads
\begin{eqnarray}
&&\Delta E(q)^2 -\Delta E(q)((1-\cos q)-k(1-\alpha))\nonumber\\
&&-\frac{k}{2}\left((1-\alpha)+k\alpha\right)(1-\cos q)=0
\label{secular}
\end{eqnarray}
and the stability condition $E(q)>0$ of the ferromagnetic phase reduces to
\begin{eqnarray}
k(\alpha-1)<1-\cos q \ \ \mbox{and} \ \ \alpha(1-k)>1
\label{ferrocond}
\end{eqnarray}
for the arbitrary $q$. This implies that the ferromagnetic ground state is stable against single-magnon excitation for
\begin{eqnarray}
\alpha >\alpha_{\rm F}\equiv\dfrac{1}{1-k}
\end{eqnarray}
for $ 0 < k <1$. The instability point $\alpha_{\rm F}$ coincides with the critical point (\ref{percr}) calculated in the limit $\alpha >> 1$. For $k >1$, the condition (\ref{ferrocond}) is never satisfied.  

The energy of the single magnon with the wave number $q$ in the neighbourhood of the instability point is given by
\begin{eqnarray}
\Delta E(q)=-\frac{k}{2}(1-(1-k)\alpha)\frac{1-\cos q}{1-\cos q +\frac{k^2}{1-k}},
\end{eqnarray}
which is negative for the arbitrary nonvanishing $q$ for $\alpha < \alpha_{\rm F}$. Therefore, all modes become unstable at $\alpha_{\rm F}$. Among them, the lowest energy mode has $q=\pi$, which corresponds to the short-range order in the Haldane phase. However, with the decrease in $k$, the $q$-dependence of $\Delta E(q)$ becomes weak and $\Delta E(q)$ tends to a $q$-independent value $-\frac{k}{2}(1-(1-k)\alpha)$. This implies that almost all modes become equally unstable at $\alpha=\alpha_{\rm F}$ for small $k$. This might be the origin of the complex feature of the phase diagram in the small $k$ regime presented in the next section.

\section{Numerical Ground State Phase Diagram}

\subsection{Ferromagnetic-nonmagnetic phase boundary}

\begin{figure}
\centerline{\includegraphics[width=70mm]{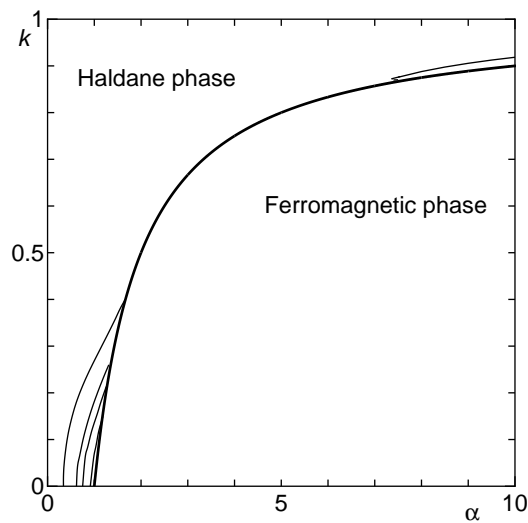}}
%\centerline{\includegraphics[width=70mm]{phaseovalext.eps}}
\caption{Overall phase diagram of  present model. The boundary of the ferromagnetic phase is denoted by a thick curve. The detailed structure of the phase diagram drawn by thin curves are explained in the text.}
\label{phase}
\end{figure}

 The overall ground state phase diagram is presented in  Fig. \ref{phase}.  The  boundary of the ferromagnetic phase is determined from the ground state level cross for the finite chains. This boundary has no system size dependence for $N=12, 16, 20$ and 24, and coincides with $\alpha_{\rm F}$ determined in the preceding section.  Therefore, the transition of the ferromagnetic phase to other phases takes place as a single magnon instability rather than a multimagnon instability. In the neighbourhood of  this phase boundary, various intermediate phases are found. The details are discussed for large $\alpha$ regime and small $k$ regime separately in the following subsections.

\subsection{Large $\alpha$ regime}

\begin{figure}
\centerline{\includegraphics[width=70mm]{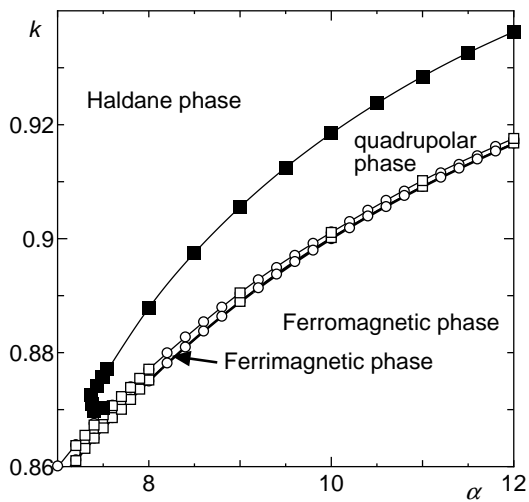}}
%\centerline{\includegraphics[width=70mm]{largekfext.eps}}
\caption{Phase diagram in large $\alpha$ regime. Filled squares indicate phase boundaries determined by the extrapolation from $N=12, 18$ and 24. Open circles and squares indicate the data for $N=12$ and $N=24$, respectively. The thick line indicates the single magnon instability point $\alpha_{\rm F}$. Thin lines are guides for eye.}
\label{largekf}
\end{figure}

As expected from the perturbational calculation in \S \ref{pert},  there exists a gapless spin-quadrupolar phase between the Haldane and ferromagnetic  phases for large $\alpha$.  This phase is identified as the nonmagnetic phase with $S=2$ gapless excitations, while the  Haldane phase is characterized by the gapped excitations with $S=1$. The phase boundary is determined from the level cross of the $S=1$ and $S=2$ excitations for $N=12, 18$ and 24 chains. The system sizes are chosen by considering the quasi-trimerized nature of the spin-quadrupolar phase\cite{ft1,ft2}.  The phase transition line extrapolated to $N \rightarrow \infty$ is presented in Fig. \ref{largekf}. Although the spin-quadrupolar phase obtained by our numerical method is quite narrow, we conclude that  this phase does exist between the spin gap phase and the ferromagnetic phase because its existence is confirmed by the perturbational calculation in \S \ref{pert}. 

\begin{figure}
\centerline{\includegraphics[width=70mm]{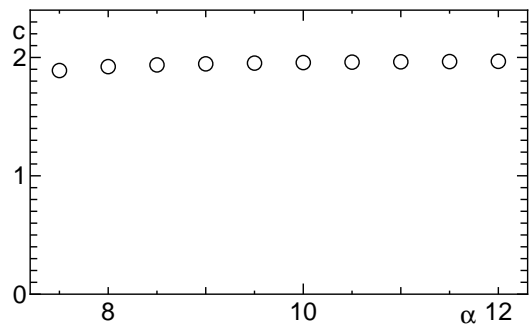}}
%\centerline{\includegraphics[width=70mm]{conf_charge_uls.eps}}
\caption{Central charge on spin-quadrupolar-Haldane phase boundary.}
\label{chargeuls}
\end{figure}

\begin{figure}
\centerline{\includegraphics[width=70mm]{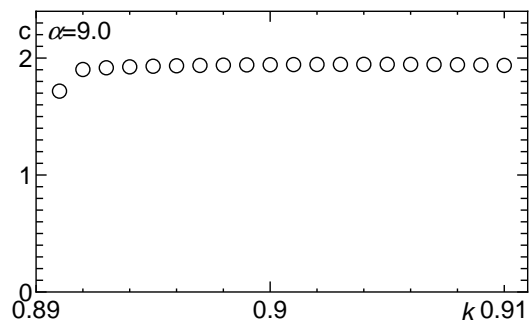}}
%\centerline{\includegraphics[width=70mm]{conf_charge_ulsncr.eps}}
\caption{Central charge within spin-quadrupolar phase for $\alpha=9.0$.}
\label{chargeulsncr}
\end{figure}

\begin{figure}
\centerline{\includegraphics[width=70mm]{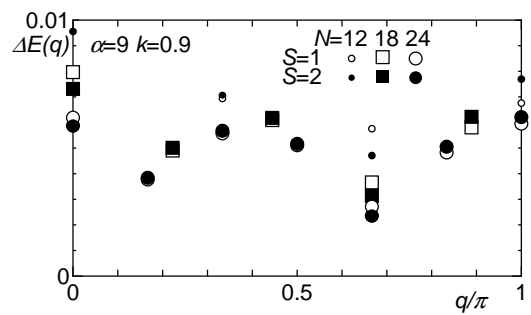}}
%\centerline{\includegraphics[width=70mm]{gapskulsk09r9s.eps}}
\caption{Dispersion relation in spin-quadrupolar phase for $\alpha=9$ and $k=0.9$ with $N=12, 18$ and 24.}
\label{dispuls}
\end{figure}

To confirm the universality class of the transition between the Haldane phase and spin-quadrupolar phase, we estimate the central charge $c$ determined by the formula
\begin{equation}
\label{car}
\frac{2}{N}E_{\mbox{g}}(N) \cong \varepsilon_{\infty}-\frac{\pi cv_{\mbox{s}}}{6(N/2)^{2}},
\end{equation}
where $v_{\rm s}$ is the spin wave velocity defined by
\begin{equation}
\label{eq2}
v_{\mbox{s}}=\lim_{N \rightarrow \infty}\frac{N}{4\pi}[E_{q_{1}}(N)-E_{\mbox{g}}(N)].
\end{equation}
Here, $E_{\mbox{g}}(N)$ is the ground state energy, $E_{q_{1}}(N)$, the excitation energy of the excited state with the wave number $q_{1}=\frac{2\pi}{(N/2)}$, and $\varepsilon_{\infty}$ is the ground state energy per unit cell in the thermodynamic limit. The extrapolation is carried out from $N=12, 18$ and 24. The results are presented in Fig. \ref{chargeuls}, which confirms $c=2$ as expected on the ULS point.\cite{ik} We have also checked $c=2$ within the spin-quadrupolar phase as shown in  Fig. \ref{chargeulsncr} at $\alpha=9$. This confirms the critical nature of this phase. It is also numerically checked that the lowest excited state has the wave number $q=2\pi/3$ and total spin $S=2$ within the spin-quadrupolar phase, as shown in Fig. \ref{dispuls}.

 Precise calculation suggests the presence of an extremely narrow ferrimagnetic region with intermediate spontaneous magnetization between the ferromagnetic phase and the nonmagnetic phase.  In Fig. \ref{largekf}, open circles are the phase boundaries estimated  for the $N=12$ chains and the open squares are those estimated for $N=24$ chains.  The system size dependence of the ferrimagnetic-nonmagnetic phase boundary is also very weak. Therefore, we expect that this ferrimagnetic phase will survive in the thermodynamic limit. 
 
 This ferrimagnetic phase can be regarded as a  frustration-induced ferrimagnetic phase, which has been found in various frustrated spin chains recently.\cite{ym,kh} This phase vanishes for sufficiently large $\alpha$ where this model reduces to the $S=1$ bilinear-biquadratic chain, which has no frustration. This phase also vanishes for small $k$, where the direct ferromagnetic-nonmagnetic transition is expected from the argument of \S \ref{bigspin}. Within the numerical data, however, it is impossible to determine whether this phase vanishes at a finite value of $\alpha$ or $k$. 
\begin{figure}
\centerline{\includegraphics[width=70mm]{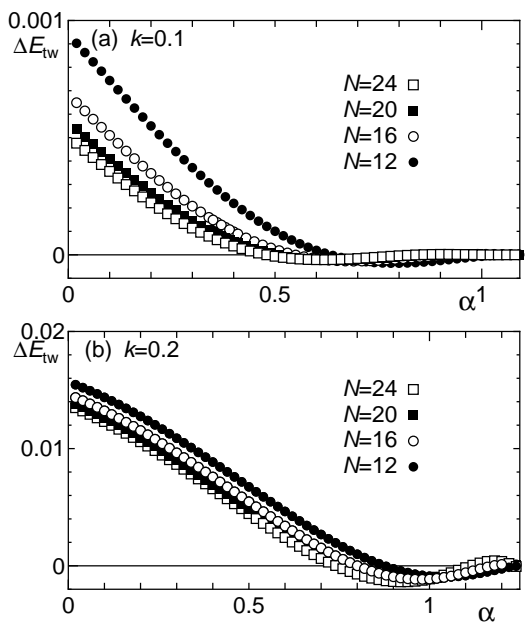}}
%\centerline{\includegraphics[width=70mm]{twkfixdiffs.eps}}
\caption{Energy difference $\Delta \Etw = E_+-E_-$ of ground states of different parities with twisted boundary condition for (a) $k=0.1$ and (b) 0.2.
}
\label{twgap}
\end{figure}
\begin{figure}
\centerline{\includegraphics[width=70mm]{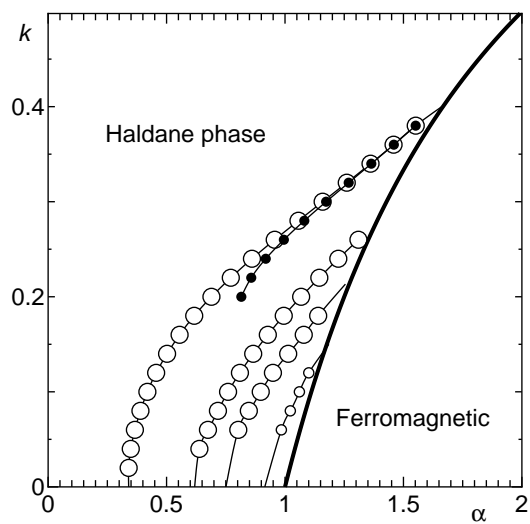}}
%\centerline{\includegraphics[width=70mm]{smallkextal.eps}}
\caption{Phase diagram in small $k$ regime. Large open circles indicate Gaussian phase boundaries determined by the twisted boundary method extrapolated from $N=16, 20 $ and 24 for $\alpha_{\rm c}^{(1)}$ and $\alpha_{\rm c}^{(2)}$, and from $N=20, 24 $ and 28 for $\alpha_{\rm c}^{(3)}$. Small open circles indicate $\alpha_{\rm c}^{(4)}$  for $N=28$. Small filled circles indicate the phase boundary determined by the phenomenological renormalization group analysis of  open chains with $12 \leq N \leq 52$, where the values of $N$ are multiples of 4. The thick solid line is $\alpha_{\rm F}$, which coincides with the numerically obtained ferromagnetic-nonmagnetic phase boundary. Thin lines are guides for eye. For very small $k$, numerical data are unavailabe owing to the poor convergence of Lanczos diagonalization.}
\label{smallk}
\end{figure}

\subsection{Small $k$ regime}

The present model has two spin-1/2 degrees of freedom in a unit cell. Therefore, it is natural to expect that this model undergoes Gaussian phase transitions between different spin gap phases as the system parameters are varied. To determine the phase transition lines, we have carried out the numerical diagonalization with twisted boundary condition and evaluated the spin inversion parity assuming the Gaussian phase transitions described by the conformal field theory with the central charge $c=1$. We twist the spins  $\v{\sigma}_2$ and $\v{\tau}_1$ by an angle $\pi$ relative to $\v{\sigma}_1$ around the $z$ axis. As discussed by Kitazawa\cite{kita} and Kitazawa and Nomura,\cite{kn} the spin inversion parity should change sign if the valence bond structure of the nonmagnetic phase changes. 
\begin{figure}
\centerline{\includegraphics[width=70mm]{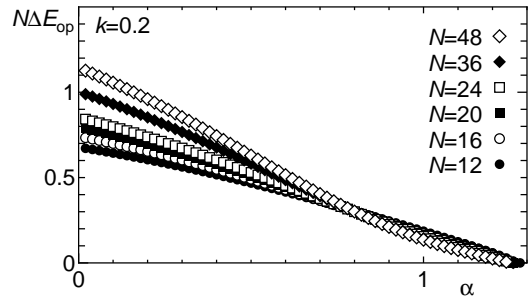}}
%\centerline{\includegraphics[width=70mm]{ogapk20s.eps}}
\caption{$\alpha$-dependence of  scaled gap with open boundary condition for  $k=0.2$.}
\label{prgop}
\end{figure}
\begin{figure}
\centerline{\includegraphics[width=70mm]{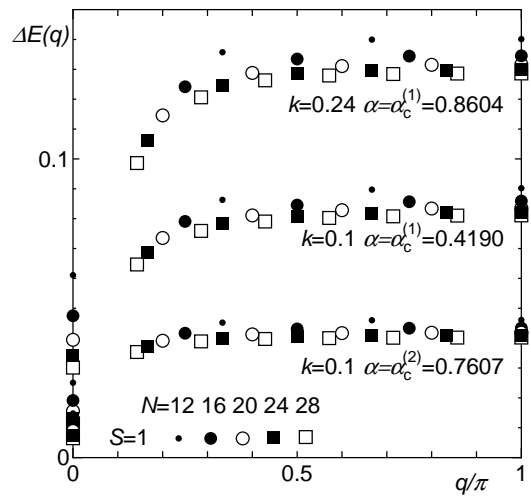}}
%\centerline{\includegraphics[width=70mm]{gapsmkcr.eps}}
\caption{Dispersion relations on critical points $\alpha_{\rm c}^{(1)}$ and $\alpha_{\rm c}^{(2)}$ for $k=0.1$ and $k=0.24$. The system sizes are $N=12, 16, 20, 24$ and 28.}
\label{gapsmkcr}
\end{figure}
\begin{figure}
\centerline{\includegraphics[width=70mm]{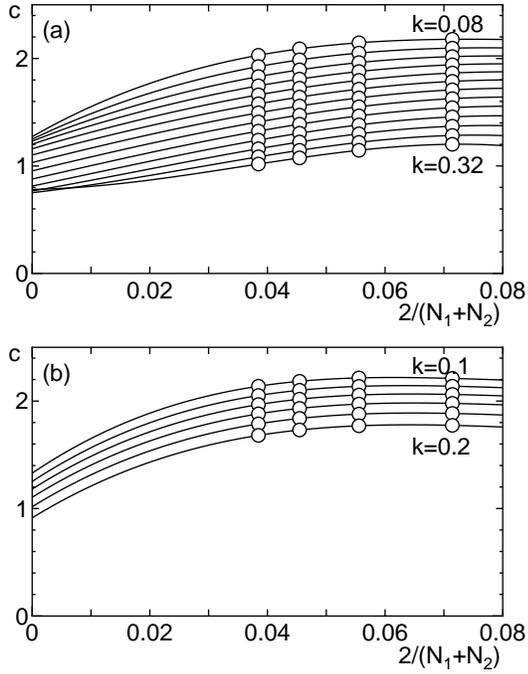}}
%\centerline{\includegraphics[width=70mm]{c_smkndep.eps}}
\caption{System size dependence of central charge estimated for finite size system for (a) $\alpha=\alpha_{c}^{(1)}$ with $0.08 \leq k \leq 0.32$ and (b)$\alpha=\alpha_{c}^{(2)}$ with $0.1 \leq k \leq 0.2$ plotted against $2/(N_1+N_2)$, where $N_1$ and $N_2$ are system sizes used for determining finite size correction to  ground state energy and spin wave velocity. The extrapolation to $(N_1+N_2)/2\rightarrow \infty$ is carried out by fitting the size dependence by the third-order polynominals. 
}
\label{chargesmk}
\end{figure}
\begin{figure}
\centerline{\includegraphics[width=70mm]{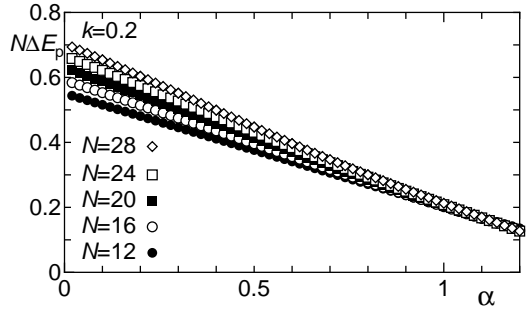}}
%\centerline{\includegraphics[width=70mm]{prgapk20as.eps}}
\caption{$\alpha$-dependence of  scaled gap $\Delta \Epr$ with periodic boundary condition for  $k=0.2$. 
}
\label{prgper}
\end{figure}
\begin{figure}
\centerline{\includegraphics[width=70mm]{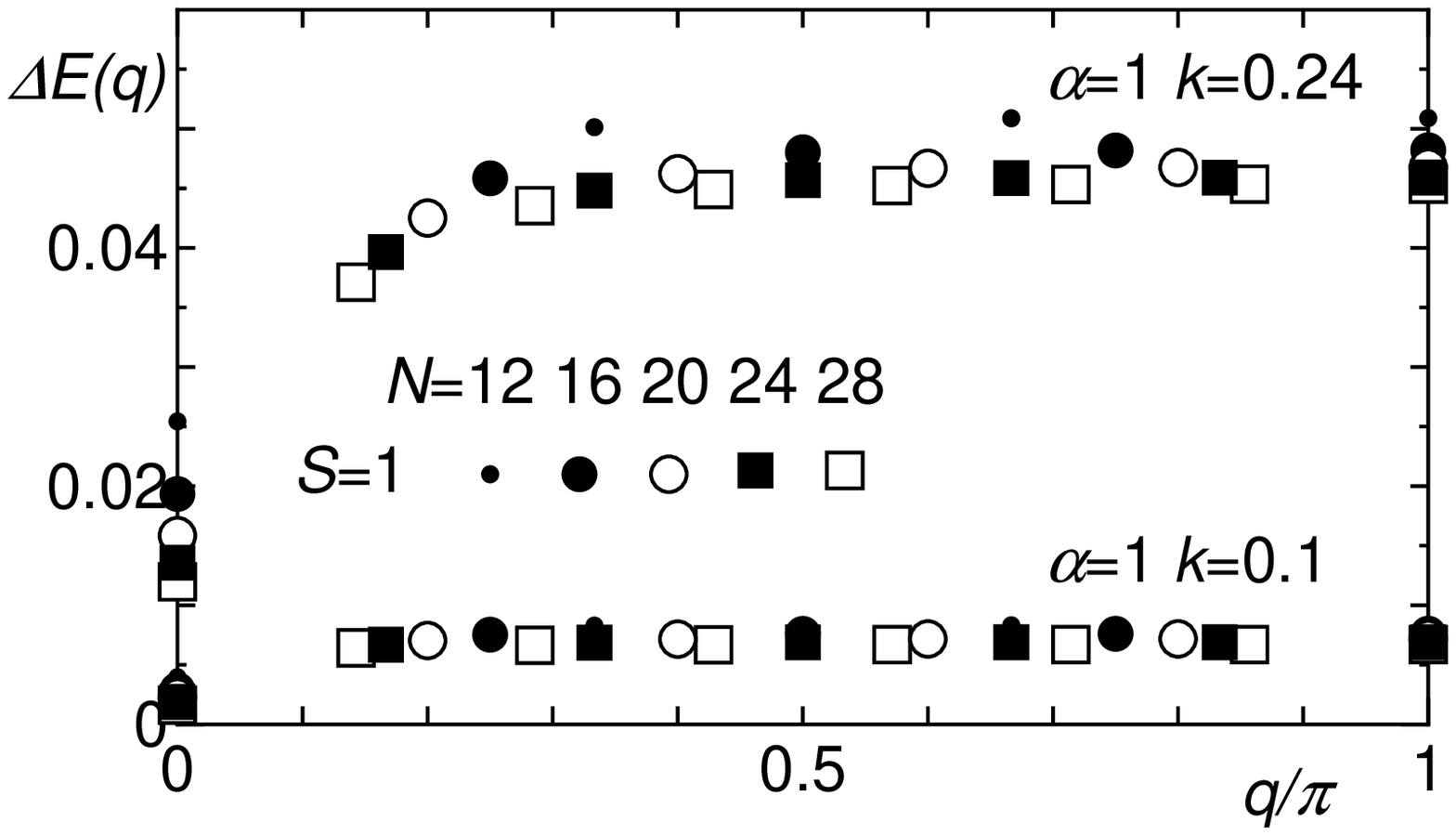}}
%\centerline{\includegraphics[width=70mm]{gapsmks.eps}}
\caption{Dispersion relation for $\alpha=1$ with $k=0.1$ and $k=0.24$. The system sizes are $N=12, 16, 20, 24$ and 28.}
\label{gapsmk}
\end{figure}

The difference $\Delta \Etw$ between the ground state energy with the positive parity $E_+$ and that with the negative parity $E_-$ is shown in Fig. \ref{twgap} for $k=0.1$ and 0.2.  Surprisingly, the ground state undergoes a series of phase transitions among the states with different parities, although the energy difference between the two ground states  is small for small $k$. The phase boundaries extrapolated to $N \rightarrow \infty$ are shown in Fig. \ref{smallk}. Let us denote the $l$-th transition point by $\alpha_{\rm c}^{(l)}(k)$ so that $\alpha_{\rm c}^{(l)}(k)<\alpha_{\rm c}^{(l+1)}(k)$. For $0 < \alpha < \alpha_{\rm c}^{(1)}(k)$, the valence bonds sit on the $K$-bonds in the ground state. In this phase, the spin inversion parity with twisted boundary condition is negative because a valence bond sits on the twisted bond. There is no phase boundary between this phase and the Haldane phase in the large $\alpha$ limit. Therefore, this phase is also called the Haldane phase in the following.

To confirm this phase transition line, we have also carried out the phenomenological renormalization group analysis\cite{domb} of the scaled singlet-triplet gap  $N\Delta \Eop$ with  open boundary condition using  exact diagonalization ($N=12,16,20$ and 24) and density matrix renormalization group ($N=36, 48$ and 52). The $\alpha$-dependence of $N\Delta \Eop$ is presented in Fig. \ref{prgop}  for $k=0.2$ and $12 \leq N \leq 48$. 
It should be noted that $N\Delta \Eop$ values for different system sizes cross with each other around the critical point  $\alpha_{\rm c}^{(1)}(k=0.2)\simeq 0.74$, suggesting that the end spins appear for $\alpha > \alpha_{\rm c}^{(1)}$. Actually, the transition points $\alpha_{\rm op}$ calculated from the crossing point of $N\Delta \Eop$ coincide well with $\alpha_{\rm c}^{(1)}$ for relatively large $k$ values ($k \simgeq 0.22$) as shown in Fig \ref{smallk} by small filled circles. For smaller values of $k$, however, they deviate from each other. The crossing points of $N\Delta \Eop$ values corresponding to $\alpha_{c}^{(l)}$ with $l \geq 2$ are not found. These discrepancies are presumably due to the logarithmic correction to $\Eop$ resulting from the SU(2) symmetry. Therefore, we rely on the result of the twisted boundary method for small $k$. Neverhteless, the coincidence for $k \simgeq 0.22$  suggests that this transition is a usual Gaussian transition between different valence bond ground states. From continuity, it is natural to expect that this will be valid even for smaller values of $k$.

The size dependence of  $\alpha_{\rm c}^{(2)}$ estimated by the twisted boundary condition method is stronger than that of  $\alpha_{\rm c}^{(1)}$. The critical line $\alpha_{\rm c}^{(3)}$  cannot be identified for chains with length $N \leq 20$ and $\alpha_{\rm c}^{(4)}$ is found only for $N=28$. These features of the spin-gap phases near the ferromagnetic phase imply that the resonation  among long ranged valence bond states\cite{th} would be required to describe these phases rather than a simple valence bond picture. This in turn suggests the possibility that  further series of transitions may take place before the ferromagnetic transition point is reached in the thermodynamic limit. 

To confirm that these critical lines are actually described by the $c=1$ conformal field theory, we have also carried out the calculation of the conformal charge $c$. Unfortunately, due to the strongly nonlinear nature of the dispersion relation  for small wave numbers shown in Fig \ref{gapsmkcr}, the extrapolated values of $c$ depend strongly on the way of extrapolation. Here, we employ the following method:  The coefficients  of eq. (\ref{car}) are determined by fitting the ground state energies of $N_1$ and $N_2(=N_1+4)$ using eq. (\ref{car}) to obtain the finite size estimation of $A(N_1,N_2)=v_{\rm s}c$. The numerical data for the excitation energy for the same set of system sizes are used to determine the finite size estimation of $v_s(N_1,N_2)$ by fitting the smallest wave number excitation energy as $\frac{N}{4\pi}[E_{q_{1}}(N)-E_{\mbox{g}}(N)]=v_s + B/N$ with the constant $B$. The finite size estimation of $c(N_1,N_2)$ is determined by $c(N_1,N_2)=A(N_1,N_2)/v_{\rm s}(N_1,N_2)$. This is extrapolated to $N_1, N_2 \rightarrow \infty$ as a function of $2/(N_1+N_2)$ as shown in Fig. \ref{chargesmk}. The values of $c$ thus extrapolated seem to converge around unity.  This result is consistent with the assumption that the present transition lines are described by the $c=1$ conformal field theory, although other possibilities cannot be totally ruled out considering the strong dependence on the system size and on the way of extrapolation.  This result justifies the use of the twisted boundary condition method to determine the phase boundary.

\section{Summary and Discussion}
The ground state phase diagram of the spin-1/2 Heisenberg $\Delta$-chain with a ferromagnetic main chain is determined both perturbationally and numerically. The overall phase diagram is divided into nonmagnetic, ferromagnetic and extremely narrow ferrimagnetic phases. For large $\alpha$, the spin-quadrupolar phase with $S=2$ gapless excitation is shown to appear as anticipated from the perturbational mapping onto the $S=1$ bilinear-biquadratic chain. It is remarkable that the spin-quadrupolar phase is found in the $S=1/2$ model with only bilinear exchange interaction. This increases the possibility of observing this phase in real materials, although the materials described by the present model is not realized so far. 

In the small $k$ regime, a series of ground states with different parities under twisted boundary condition are found  near the nonmagnetic-ferromagnetic phase boundary. It is assumed that these phases are stabilized by the resonation of various long-distance valence bond configurations. However, the ground state of this parameter regime is not fully understood within the present numerical studies. Here we present some arguments towards deeper understanding of phases in this regime. 

The energy difference $\Delta\Etw$ between the ground states with different parities is extremely small  for $\alpha_{\rm F} >\alpha > \alpha_{\rm c}^{(2)}(k)$ with small $k$, as shown in Fig. \ref{twgap}. The phenomenological renormalization group analysis  of $\Delta \Eop$ fails to find $\alpha_{c}^{(l)}$'s with $l \geq 2$, as shown in Fig. \ref{prgop}. The gap with the periodic boundary condition is almost critical in this regime as shown in Fig. \ref{prgper}. From these behaviors, one might be tempted to conclude that the ground state in this regime is actually gapless and Gaussian lines obtained here are stable fixed lines, that do not correspond to phase transitions. However, such an extended gapless phase with triplet lowest excitation is  unlikely in the present model because the gap formation term due to period-2 spatial modulation is always relevant at the SU(2) invariant point of $c=1$ conformal field theory. The spin quadrupolar phase is also excluded because it  should have spin-2 lowest excitation. In addition, no trace of quasi-trimerization is found in the dispersion relation within this phase as shown in Fig. \ref{gapsmk}.

Therefore, we may speculate that the phases for $\alpha_{\rm F} >\alpha > \alpha_{\rm c}^{(2)}(k)$ are true gapped phases  with an extremely small gap. A similar phase is found  in  the spin-1/2 zigzag chain with ferromagnetic nearest neighbour and antiferromagnetic next nearest neighbour couplings,\cite{iq} which  has been fairly well investigated as a one-dimensional example of nonmagnetic phase that emerge from ferromagnetism.\cite{hamada,tonehara,iq} However, the nonmagnetic ground state of this model has an incommensurate spin correlation and ground state degeneracy in the thermodynamic limit accompanied by spontaneous dimerization\cite{tonehara,iq}. In contrast, in the present $\Delta$-chain model with small $k$, no signal of ground state degeneracy is found and the dispersion relation has no structure at any incommensurate wave numbers, as shown in Figs. \ref{gapsmkcr} and \ref{gapsmk}. Presumably, this is because the antiferromagnetic bonds in the present model do not compete with each other in forming valence bonds. Thus, we speculate that the nonmagnetic ground state of the present model with small $k$ and $\alpha \simleq \alpha_{\rm F}$ can be regarded as a new nonmagnetic state with a highly reduced gap realized by the influence of frustrated coupling to the apical spins. Unfortunately, we do not find an appropriate low-energy effective field theory  with a markedly reduced mass that describes this phase. Therefore,  full understanding of this ground state is left for future studies.

It would be an interesting experimental challenge to synthesize the material corresponding to the present model and observe the exotic nonmagnetic phases described in this paper. Considering that the materials with a $\Delta$-chain geometry are already synthesized,\cite{inagaki,sswc,tend} it is promising to expect the experimental realization of the present model in the near future.

The author is grateful to K, Takano for valuable discussion. The numerical diagonalization program is based on the package TITPACK ver. 2 coded by H. Nishimori.  The numerical computation in this work has been carried out using the facilities of the Supercomputer Center, Institute for Solid State Physics, University of Tokyo and Supercomputing Division, Information Technology Center, University of Tokyo. This work is partly supported by Innovative Research Organization, Saitama University.


\begin{thebibliography}{10}
\bibitem{mg} C. K. Majumdar and D. K. Ghosh: J. Math. Phys. {\bf 10} (1969) 1399.
\bibitem{oku} K. Okunishi and T. Tonegawa: J. Phys. Soc. Jpn. {\bf 72} (2003) 479.
\bibitem{oku2} K. Okunishi and T. Tonegawa: Phys. Rev. B{\bf 68} (2003) 224422.
\bibitem{tone} T. Tonegawa, K. Okamoto, K. Okunishi, K. Nomura and M. Kaburagi: Physica {B} {\bf 346-347} (2004) 50.
\bibitem{ha} K. Hida and I. Affleck:   J. Phys. Soc. Jpn. {\bf 74} (2005)  1849.
\bibitem{ym} S. Yoshikawa and S. Miyashita:   J. Phys. Soc.Jpn. {\bf 74} Suppl.(2005) 71.
\bibitem{kh} K. Hida: J. Phys.: Condens. Matter 19 (2007) 145225.
%delta chains
\bibitem{hamada} T. Hamada, J. Kane, S. Nakagawa and Y. Natsume: J. Phys. Soc. Jpn.
{\bf 57} (1988) 1891.
\bibitem{kubo} K. Kubo: Phys. Rev. B{\bf 48} (1993) 10552.
\bibitem{nk} T. Nakamura and K. Kubo: Phys. Rev. B{\bf 53} (1996) 6393.
\bibitem{tonedelta} T. Tonegawa and M. Kaburagi: J. Magn. Magn. Mater. {\bf 272-276}(2004) 898.
\bibitem{sswc} D. Sen, B. S. Shastry, R. E. Walstedt and R. Cava: Phys. Rev. B{\bf 53} (1996) 6401.
\bibitem{inagaki} Y. Inagaki, Y. Narumi, K. Kindo, H. Kikuchi,
T. Kamikawa, T. Kunimoto, S. Okubo, H. Ohta, T. Saito, M. Azuma, M. Takano, H. Nojiri, M. Kaburagi and T. Tonegawa: J. Phys. Soc. Jpn. {\bf 74} (2005) 2831.
\bibitem{tend} G. Van Tendeloo, O. Garlea, C. Darie, C. Bougerol-Chaillout and P. Bordet: J. Solid State Chem. {\bf 156} (2001) 428.
\bibitem{khfa} K. Hida: Phys. Rev. B{\bf 46}  (1992) 2207.

\bibitem{th} K. Takano and K. Hida:  arXiv:0710.3457.
% uls phases
\bibitem{uimin} G. V. Uimin: JETP Lett. {\bf 12} (1970) 225.
\bibitem{lai} C. K. Lai: J. Math. Phys. {\bf 15} (1974) 1675. 
\bibitem{suth} B. Sutherland: Phys. Rev. B{\bf 12} (1975) 3795.
\bibitem{ft1} G. F\'ath and J. S\'olyom: Phys. Rev. B{44} (1991) 11836.
\bibitem{ft2} G. F\'ath and J. S\'olyom: Phys. Rev. B{47} (1993) 872.
\bibitem{ik} C. Itoi and M. Kato: Phys. Rev. B{\bf 55} (1997) 8295.
\bibitem{lauchli} A. Lauchli,G. Schmid and S. Trebst: Phys. Rev. B{\bf 74} (2006) 144426.
% ferro leg ladders
\bibitem{doniach} S. Doniach: Physica B{\bf 91} (1977)231.
\bibitem{sss} R. T. Scalettar, D. J. Scalapino and R. L. Sugar: Phys. Rev. B{\bf 31}(1985) 7316.
\bibitem{on} H. Otsuka and T. Nishino:  Phys. Rev. B{\bf 52}7316 (1995) 15066.
\bibitem{yama1} T. Yamamoto, M. Asano and C. Ishii: J. Phys. Soc. Jpn. {\bf 70} (2001) 3678.
\bibitem{yama2} T. Yamamoto, K. Ide and C. Ishii: Phys. Rev. B{\bf 66} (2002) 104408.
\bibitem{yama3}T. Yamamoto, R. Manago, Y. Mori and C. Ishii: J. Phys. Soc. Jpn. {\bf 72} (2003) 3204.

\bibitem{km1} A. K. Kolezhuk and H.-J. Mikeska: Phys. Rev. B{\bf 53} (1996) R8848.
\bibitem{vjm}T. Vekua, G. I. Japaridze and H.-J. Mikeska: Phys. Rev. B{\bf 67}  (2003) 064419.
\bibitem{rm} M. Roji and S. Miyashita:  J. Phys. Soc. Jpn. {\bf 65} (1996) 883.
\bibitem{hijii} K. Hijii, K. Nomura, and A. Kitazawa: Phys. Rev. B{\bf 72} (2005) 014449.
\bibitem{hase} M. Hase, H. Kuroe, K. Ozawa, O. Suzuki, H. Kitazawa, G. Kido and
T. Sekine: Phys. Rev. B{\bf 70} (2004) 104426.
\bibitem{kage} H. Kageyama, T. Kitano, N. Oba, M. Nishi, S. Nagai, K. Hirota, L. Viciu, J. B. Wiley, J. Yasuda, Y. Baba, Y. Ajiro and K. Yoshimura: J. Phys. Soc. Jpn. {\bf 74} (2005) 1702.
\bibitem{tonehara} T. Tonegawa and I. Harada: J. Phys. Soc. Jpn. {\bf 56} (1987) 2153.
\bibitem{iq} C. Itoi and S. Qin: Phys. Rev. B{\bf 63} (2001) 224423.
\bibitem{momoi} N. Shannon, T. Momoi and P. Sindzingre: Phys. Rev. Lett. {\bf 96} (2006) 027213.
\bibitem{kita} A. Kitazawa: J. Phys. A Math. Gen. {\bf 30} (1997) L285.
\bibitem{kn} A. Kitazawa and K. Nomura: J. Phys. Soc. Jpn. {\bf 66} (1997) 3944.
\bibitem{domb} M. N. Barber: in {\it Phase Transition and Critical Phenomena,} ed. C. Domb and J. L. Lebowitz, (Academic Press, London, New York, 1983) Vol. 8.
\end{thebibliography}
\end{document}